\documentclass[12pt,preprint]{aastex}

\bibpunct[,]{(}{)}{;}{a}{}{,}

\begin{document}


\def\msun{$M_{\odot}$}
\def\rsun{$R_{\odot}$}
\def\mdot{$\dot M$}
\def\eddmdot{$\dot m$}
\def\RS{$R_{\rm S}$}
\def\Rg{$R_{\rm g}$}
\def\fdg{\hbox{$.\!\!^\circ$}}
\def\fas{\hbox{$.\!\!\*(U$}}
\def\fss{\hbox{$.\!\!^s$}}
\def\ergsec{\hbox{erg s$^{-1}$ }}
\def\ergseccm{\hbox{erg s$^{-1} $cm$^{2}$ }}
\def\ergcm{\hbox{erg cm$^{-2}$ s$^{-1}$ }}
\def\erga{\hbox{erg cm$^{-2}$ s$^{-1}$ \AA$^{-1}$ }}
\def\ergh{\hbox{erg cm$^{-2}$ s$^{-1}$ Hz$^{-1}$ }}
\def\fcol{$f_{\rm col}$}
\def\Tin{$T_{\rm in}$}
\def\Teff{$T_{\rm eff}$}

\title{Measuring the Spins of Accreting Black Holes}

\author{Jeffrey E. McClintock\altaffilmark{1}, 
  Ramesh Narayan\altaffilmark{1},
  Shane W. Davis\altaffilmark{2},
  Lijun Gou\altaffilmark{1},\ 
  Akshay Kulkarni\altaffilmark{1}, 
  Jerome A. Orosz\altaffilmark{3},
  Robert F. Penna\altaffilmark{1},
  Ronald A. Remillard\altaffilmark{4}, 
  James F. Steiner\altaffilmark{1}}

\altaffiltext{1}{Harvard-Smithsonian Center for Astrophysics, 60
  Garden Street, Cambridge, MA 02138}
\altaffiltext{2}{Canadian Institute for Theoretical
Astrophysics, Toronto, ON M5S3H4, Canada}
\altaffiltext{3}{Department of Astronomy, San Diego State University, 5500
  Companile Drive, San Diego, CA 92182}
\altaffiltext{4}{MIT Kavli Institute for Astrophysics and Space
  Research, MIT, 70 Vassar Street, Cambridge, MA 02139}

\begin{abstract}

A typical galaxy is thought to contain tens of millions of
stellar-mass black holes, the collapsed remnants of once massive
stars, and a single nuclear supermassive black hole.  Both classes of
black holes accrete gas from their environments. The accreting gas
forms a flattened orbiting structure known as an accretion disk.
During the past several years, it has become possible to obtain
measurements of the spins of the two classes of black holes by
modeling the X-ray emission from their accretion disks.  Two methods
are employed, both of which depend upon identifying the inner radius
of the accretion disk with the innermost stable circular orbit (ISCO),
whose radius depends only on the mass and spin of the black hole.  In
the Fe K$\alpha$ method, which applies to both classes of black holes,
one models the profile of the relativistically-broadened iron line
with a special focus on the gravitationally redshifted red wing of the
line.  In the continuum-fitting method, which has so far only been
applied to stellar-mass black holes, one models the thermal X-ray
continuum spectrum of the accretion disk.  We discuss both methods,
with a strong emphasis on the continuum-fitting method and its
application to stellar-mass black holes.  Spin results for eight
stellar-mass black holes are summarized.  These data are used to argue
that the high spins of at least some of these black holes are natal,
and that the presence or absence of relativistic jets in accreting
black holes is not entirely determined by the spin of the black hole.
\end{abstract}

\section{Introduction}

Our focus throughout is on the two principal methods for measuring the
spins of accreting black holes: modeling the thermal continuum X-ray
spectrum and modeling the profile of the relativistically-broadened Fe
K$\alpha$ line.  The continuum-fitting (CF) method, has thus far only
been applied to stellar-mass black holes in X-ray binaries, whereas
the Fe K$\alpha$ method has been applied to both stellar-mass ($M\sim
10$~\msun) and supermassive ($M\sim 10^6-10^{10}$~\msun) black
holes\footnote{1 \msun~= 1 solar mass = $2.0 \times 10^{33}$~g.}.
This paper is chiefly focused on measuring the spins of stellar-mass
black holes via the CF method because we have been deeply engaged in
this work during the past several years.  In Section 6, we secondarily
discuss the Fe K$\alpha$ method, which is very important because it is
the primary approach to measuring the spins of supermassive black
holes.

We note that spin estimates have also been obtained by modeling the
high-frequency X-ray oscillations (100--450~Hz) observed for several
stellar-mass black holes \citep{wag+2001,tor+2005}.  At present, this
method is not providing dependable results because the correct model
of these oscillations is not known.  X-ray polarimetry is another
potential avenue for measuring spin \citep{dov+2008,lil+2009,sch+2009},
which may be realized soon with the 2014 launch of the Gravity and
Extreme Magnetism Small Explorer \citep{swa+2008}.  Meanwhile, several
other methods of measuring spin have been proposed or applied (e.g.,
\citealp{tak+2004,bar+2004,hua+2007,sul+2008,shc+2011}).


\section{Stellar-Mass Black Holes in X-ray Binaries}

Observations in 1972 of the X-ray binary Cygnus X-1 provided the first
strong evidence that black holes exist.  Today, a total of 23 such
X-ray binary systems are known that contain a compact object too
massive to be a neutron star or a degenerate star of any kind (i.e.,
$M>3$~\msun; \citealt{oze+2010}).  These compact objects, which have
typical masses of $10$\msun, are referred to as black holes.  Their
host systems are mass-exchange binaries containing a nondegenerate
star that supplies gas to the black hole via a stellar wind or via
Roche-lobe overflow in a stream that emanates from the inner
Lagrangian point.  The mass-donor star in the Roche-lobe overflow
systems is typically a low mass ($M\sim1$~\msun) sun-like star, and
the X-ray source is transient, alternating between yearlong outbursts
($L_{\rm max} \sim L_{\rm Edd} =
1.3{\times}10^{39}{M/10M_{\odot}}$~erg s$^{-1}$) and years or decades
of quiescence ($L \sim 10^{-7}L_{\rm Edd}$)\footnote{$L_{\rm Edd}$ is
the critical Eddington luminosity above which radiation pressure
exceeds gravity.}.  The wind-fed X-ray sources, on the other hand, are
fueled by massive hot stars ($M\gtrsim10 M_{\odot}$) and are
persistently luminous.  A schematic sketch to scale of 21 of these
systems is shown in Figure~1: The four at top are persistent systems
and the 17 at bottom are the transients.  For a review of the
properties of black hole binaries, see \citet{rem+2006}.

\section{The Continuum-Fitting Method}

A definite prediction of relativity theory is the existence of an
innermost stable circular orbit (ISCO) for a particle orbiting a black
hole. This inherently relativistic effect has a major impact on the
structure of an accretion disk. At radii $R \geq R_{\rm ISCO}$ (the
radius of the ISCO), accreting gas moves on nearly circular orbits and
slowly spirals in toward the black hole. At the ISCO, however, the
dynamics change suddenly and the gas, finding no more stable circular
orbits, plunges into the hole.  In the continuum-fitting (CF) method,
one identifies the inner edge of the accretion disk with the ISCO (see
Secs.\ 4 and 5 for supporting evidence) and estimates $R_{\rm ISCO}$
by fitting the X-ray continuum spectrum.  Since the dimensionless
radius $r_{\rm ISCO} \equiv R_{\rm ISCO}/(GM/c^2)$ is solely a
monotonic function of the black hole spin parameter $a_*$ (Fig.\
2)\footnote{We express black hole spin in the customary way as the
dimensionless quantity $a_* \equiv cJ/GM^2$ with $|a_*| \le 1$, where
$M$ and $J$ are respectively the black hole mass and angular
momentum.}, knowing its value allows one immediately to infer the
value of $a_*$.  We note that the truncation of the disk at the ISCO
is also a crucial assumption of the Fe K$\alpha$ method of measuring
spin \citep{rey+2008}.

Before describing the CF methodology, we stress that for this
technique to succeed it is essential to have accurate measurements of
the distance $D$ to the source, the inclination $i$ of the accretion
disk, and the mass $M$ of the black hole (for reasons discussed
below).  The methodologies for measuring $D$, $i$ and $M$ are firmly
established.  Therefore, rather than digressing to discuss how these
measurements are made, we refer the interested reader to some recent
papers on the subject \citep{oro+2007, oro+2009, oro+2011, can+2010}.

The gaseous matter flowing from the companion star to the black hole
has appreciable angular momentum as a consequence of the binary
orbital motion.  As the gas flows, viscous forces cause it to spread
out into an orbiting structure known as an accretion disk.  The gas
flowing into the outer disk spirals slowly inward on Keplerian orbits
on a time scale of weeks, reaching a typical temperature near the ISCO
of $kT\sim1$~keV.  Because the accretion disk is of fundamental
importance to the measurement of black hole spin, we now describe in
some detail the thin-disk model we employ.

The model we use is that described in \citet[][~hereafter
NT]{nov+1973}, which is a relativisitic generalization of a Newtonian
model developed by \citet{sha+1973}.  The NT model describes an
axisymmetric radiatively-efficient accretion flow which, for a given
black hole mass $M$, mass accretion rate $\dot M$ and black hole spin
parameter $a_*$, gives a precise prediction for the local radiative
flux $F(R)$ emitted at each radius $R$ of the disk.
Moreover, the accreting gas is optically thick and the emission is
thermal and blackbody-like, making it straightforward to compute the
spectrum of the emission.  Most importantly, the inner edge of the
disk is located at the ISCO.  Therefore, from the measurement of
$R_{\rm ISCO}$, and if we know the mass $M$ of the black hole, we can
immediately obtain $a_*$ (Fig.\ 2).  This is the principle behind the
CF method of estimating black hole spin, which was first described by
\citet{zha+1997}.

Before discussing how to measure $R_{\rm ISCO}$ of a disk, we remind
the reader how one measures the radius $R_*$ of a star.  Given the
distance $D$ to the star, the radiation flux $F_{\rm obs}$ received
from the star, and the temperature $T$ of its continuum radiation, 
the luminosity of the star is given by
\begin{equation}
L_* = 4\pi D^2 F_{\rm obs} = 4\pi R_*^2 \sigma T^4,
\end{equation}
where  $\sigma$ is the
Stefan-Boltzmann constant.  From $F_{\rm obs}$ and $T$, we immediately
obtain the quantity $\pi (R_*/D)^2$, which is the solid angle
subtended by the star.  From this and the distance $D$, we immediately
obtain $R_*$.  For accurate results we must allow for limb darkening
and other non-blackbody effects in the stellar emission by computing a
stellar atmosphere model, but this is a detail.

The same principle applies to an accretion disk, but with some
differences.  First, since the flux $F(R)$ emitted locally by the disk
varies with radius $R$, the radiation temperature $T(R)$ also varies
with $R$.  But the precise variation is known (if we assume the NT
disk model), so it is easily incorporated into the model.  Second,
since the bulk of the emission is from the very inner regions of the
disk, the effective area of the radiating surface is directly
proportional to the square of the disk inner radius, $A_{\rm eff} = C
R_{\rm ISCO}^2$, where the constant $C$ is known.  Third, the observed
flux $F_{\rm obs}$ depends not only on the luminosity and the
distance, but also on the inclination $i$ of the disk to the line of
sight.  Allowing for these differences, one can write a relation for
the disk problem similar in spirit to eq.\ (1), i.e., given $F_{\rm
obs}$ and a characteristic $T$ (from X-ray observations), one obtains
the solid angle subtended by the ISCO: $\pi \cos i\, (R_{\rm
ISCO}/D)^2$.  If we further know $i$ and $D$, we obtain $R_{\rm ISCO}$; and if
we also know $M$, we obtain $a_*$ (Fig.\ 2).  This is the basic idea
of the CF method.

There are three main issues that must be dealt with before applying
the method: (1) One must carefully trace rays from the disk to the
observer in the Kerr metric of the rotating black hole in order to
compute accurately the observed flux and spectrum.  To this end, we
have developed an accretion disk model called {\sc kerrbb}\footnote
{This model name and those that follow designate publicly-available
programs that comprise a suite of X-ray data analysis software known
as XSPEC (\citealt{arn+1996};
http://heasarc.gsfc.nasa.gov/docs/xanadu/xspec/index.html).}
\citep{lil+2005} for fitting X-ray data. {\sc kerrbb} assumes the NT
model of the disk and carries out all the necessary ray-tracing to
relate disk properties to observables. (2) One must have an accurate
model for computing the spectral hardening factor $f = T/T_{\rm eff}$,
where $T$ is the temperature of the radiation at a given radius and
$T_{\rm eff}$ is the effective temperature at the same radius defined
by $F(R)=\sigma T_{\rm eff}^4(R)$.  This correction for non-blackbody
effects is important at the high temperatures typically found in black
hole disks.  To carry out this correction we use the advanced disk
atmosphere models of \citet{dav+2005} and \citet{dav+2006}.  (3) Most
importantly, the inner accretion disk must be well described by the
standard geometrically-thin, optically-thick NT disk model that we
employ.  To ensure this, we restrict our attention strictly to
observations with a strong thermal component \citep{ste+2009a} and
with disk luminosities below 30\% of the Eddington limit
\citep{mcc+2006}.  In the two sections that follow, we present
observational and theoretical evidence that at these luminosities the
NT model is quite accurate.

For a full description of the mechanics of the CF method we refer
the reader to Section~4 in \citet{mcc+2006}.  In brief, we fit the
broadband X-ray continuum spectrum in conjunction with other
components as needed, principally a Compton tail component that in
more recent work is described using an empirical model of
Comptonization called {\sc simpl} \citep{ste+2009b}.  The
accretion-disk component, which is key for the CF method, is modeled
using {\sc kerrbb} \citep{lil+2005}, which includes all relativistic
effects within the context of the NT model, and also incorporates the
advanced treatment of spectral hardening mentioned above.  It
furthermore includes self-irradiation of the disk (``returning
radiation'') and the effects of limb darkening.  The key fit
parameters returned are the black hole spin parameter $a_*$ and the
mass accretion rate \mdot.


\section{Truncation of the Accretion Disk at the ISCO: Observational Evidence}

A crucial assumption that underlies both the CF and Fe~K$\alpha$
approaches to measuring spin is that the accretion disk is quite
sharply truncated at the ISCO.  This assumption, which is fundamental
to the NT model, is clearly valid if one considers only geodesic
forces in the midplane.  However, there are strong magnetohydrodynamic
(MHD) forces acting in black hole accretion disks, and it is therefore
unclear a priori that the disk terminates sharply at the ISCO.  In
this section, we present the observational evidence that there exists
a fixed inner-disk radius in black hole binaries and, in the following
section, we discuss the theoretical evidence for identifying this
radius with $r_{\rm ISCO}$.

There is a long history of evidence suggesting that fitting the X-ray
continuum is a promising approach to measuring black hole spin.  This
history begins in the mid-1980s with the application of a simple
nonrelativistic multicolor disk model \citep{mit+1984,mak+1986}, now
known as {\sc diskbb}, which returns the color temperature $T_{\rm
in}$ at the inner-disk radius $R_{\rm in}$.  In their review paper on
black hole binaries, \citet{tan+1995} summarize examples of the steady
decay (by factors of 10--100) of the thermal flux of transient sources
during which $R_{\rm in}$ remains quite constant (see their Fig.\
3.14).  They remark that the constancy of $R_{\rm in}$ suggests that
this fit parameter is related to the radius of the ISCO.
\citet{zha+1997} then outlined how, using a relativistic disk model
and corrections for the effects of radiative transfer, the fixed inner
disk radius provides an observational basis to infer black hole spin.
More recently, the evidence for a constant inner radius in the thermal
state has been presented for a number of sources via plots showing
that the bolometric luminosity of the thermal component is
approximately proportional to $T_{\rm in}^4$
\citep[e.g.,][]{kub+2001,kub+2004,gie+2004,abe+2005,mcc+2009,dun+2010}.

A recent study of the persistent source LMC X-3 presents the most
compelling evidence to date for a constant inner-disk radius
\citep{ste+2010a}.  We analyzed many spectra collected during eight
X-ray missions that span 26 years.  As shown in Figure~3, for a
selected sample of hundreds of spectra obtained using the {\it Rossi
X-ray Timing Explorer} ({\it RXTE}), we find that to within
$\approx2$~percent the inner radius of the accretion disk is constant
over time and unaffected by the gross variability of the source (top
panel).  Meanwhile, even considering an ensemble of eight X-ray
missions, we find consistent values of the radius to within
$\approx4$~percent.  These results provide compelling evidence for the
existence of a fixed inner-disk radius and establish a firm foundation
for the measurement of black hole spin.  The only reasonable inference
is that this radius is closely associated with the radius of the ISCO,
as we show to be the case in the following section.

\section{Truncation of the Accretion Disk at the ISCO: Theoretical Evidence}

The relativistic NT model on which the CF method is currently built
assumes, as does its predecessor the Newtonian model of Shakura \&
Sunyaev (1973), that the accretion disk under consideration is
geometrically thin. That is, the model assumes that the vertical
thickness $H$ at any radius $R$ satisfies $H\ll R$. Assuming in
addition that the disk is axisymmetric and in steady state, the model
derives a number of relations which follow directly from basic
conservation laws. One of the powerful results of this analysis is a
formula for the disk flux profile $F(R)$ which depends only on the
mass $M$ and spin $a_*$ of the black hole and the mass accretion rate
$\dot{M}$, but is independent of messy details such as the viscosity
of the accreting gas.  It is the existence of this robust result for
$F(R)$ that enables the CF method to work so well.

There is, however, one unproven assumption in the NT model which is
incorporated via a boundary condition: The model assumes that the
shear stress (which drives the accretion at larger radii) vanishes at
the ISCO. This ``zero-torque'' assumption is intuitively reasonable
(since the gas switches to a rapidly plunging state once it crosses
the ISCO, why should there be a stress at the transition radius?), but
as NT themselves realized, it is ultimately an assumption.
\citet{pac+2000} and \citet{afs+2003} argued that deviations from the
NT model decrease monotonically with decreasing disk thickness and
that thin disks with $H/R\ll1$ should be very well described by the
model. Their argument, which was based on a hydrodynamical description
of the disk, was confirmed by detailed calculations carried out by
\citet{sha+2008b}.  However, this leaves open the question of whether
{\it magnetized} disks might deviate substantially from the NT model.
In their paper, NT explicitly mention that magnetized disks could very
well violate the zero-torque boundary condition.

Arguments have been advanced to suggest that a magnetized accreting
gas will indeed have a non-zero shear stress at the ISCO
\citep{kro+1999,gam+1999}, and that furthermore this stress could be
so large that it may completely invalidate the NT model even in very
thin disks.  This is clearly an important question that strikes at the
heart of the CF method.  A number of recent studies of magnetized
disks using three-dimensional general relativistic magnetohydrodynamic
(GRMHD) simulations have explored this question
\citep{sha+2008a,rey+2008,nob+2009,nob+2010,pen+2010}.  The conclusion
of these authors is that the shear stress and the luminosity of the
simulated disks do differ from the NT model, but perhaps not by a
large amount.

Figure 4, taken from \citet{kul+2010}, shows the disk luminosity
distribution $dL/d\ln R =4\pi R^2 F(R)$ derived from a set of four
GRMHD thin-disk models simulated by \citet{pen+2010}. These models
have thicknesses $H/R\sim 0.045-0.08$ (see \citeauthor{pen+2010}~and
\citeauthor{kul+2010}~for a precise definition of $H$, which varies
directly with luminosity).  As is clear from the figure, the numerical
models follow the NT model reasonably well, although they do deviate
from it.  Two kinds of deviation are seen. First, the numerical models
produce some radiation inside the ISCO, whereas the NT model predicts
no radiation there. Second, the peak of the emission in the simulated
disks is shifted inward relative to the peak in the NT model.  Both of
these effects cause the disk to appear to have a smaller ISCO
radius. This in turn means that, if one were to infer the black hole
spin by fitting this luminosity distribution (or the corresponding
spectrum) using the NT model, one would infer an erroneously large
value for the spin. This systematic error arises because the NT model
is not a perfect description of the simulated disk. To the extent that
the simulated disk is a closer match to a real accretion disk than is
the NT model, this allows us to estimate the corresponding systematic
error in our measurements of spin.

How serious is this systematic error? We answer this in three parts
\citep[see][~for details]{kul+2010}.

1. Each of the models shown in Figure~4 causes an error in the spin
estimate that is smallest for a low disk inclination (face-on disk;
see Fig.\ 1) and largest for a high disk inclination ($i=75^{\circ}$).
For the latter (most unfavorable) case, the four models, which
correspond to true spin values of $a_*=0$, 0.7, 0.9 and 0.98, give via
the CF method spin values of 0.17, 0.83, 0.936 and 0.991,
respectively.

2. A similar exercise can be carried out for disks with other
thicknesses. It is found that the error in the spin estimate is larger
for thicker disks and smaller for thinner disks. Very roughly, the
error appears to scale as $H/R$. Thus, the conclusion
of \citet{pac+2000}, \citet{afs+2003} and \citet{sha+2008b},
that deviations from the NT model vanish in the limit of
vanishingly small disk thickness, appears to be valid also for MHD
disks.

3. The particular models shown in Figure~4 correspond to disk
luminosities in the range $L/L_{\rm Edd} \sim 0.4-0.8$, based on
their $H/R$. (It is difficult to be very quantitative since the
mapping between luminosity and $H/R$ is not known precisely.)  Since
the CF method is applied only to observations at $L/L_{\rm Edd} <0.3$,
the systematic error due to inaccuracies in the theoretical model
could be up to a factor of 2 smaller than the errors quoted in point 1
above.

Although the above results are based on numerical simulations that do
not necessarily mimic real disks perfectly, we believe they still
provide an estimate of the likely magnitude of the systematic error.
The key point is that the level of systematic error we find is not
serious at the current time.  The observational errors considered in
the following section are significantly larger in all cases.  Stating
this differently, while magnetized disks do behave as if their inner
edges are shifted inward relative to the position of the ISCO
(Fig.~4), the effect is quantitatively not serious for the disk
luminosities (or disk thicknesses) at which the CF method is applied.


\section{Results of Continuum Fitting}

The spin results obtained to date for eight stellar-mass black holes
are summarized in Table~1.  The spins we find cover the full allowable
range of prograde spins (Fig.~2) from $a_*\approx0$ (Schwarzschild) to
$a_*\approx1$ (extreme Kerr).  Interestingly, the spin values that
have been obtained (Table~1), while spanning the full range of
prograde spins, are all in the canonical physical range, namely $|a_*|
\le 1$.  This is an important result and not at all a foregone
conclusion: Given the hard external constraints on the dynamical model
parameters ($D$, $i$ and $M$; Section~3), it is entirely possible that
a black hole will be found that implies a spin beyond the reach of our
current model, i.e., $|a_*|>1$.  Such a result could simply be caused
by large systematic errors in $D$, $i$ and $M$.  Or, more
interestingly, it could falsify our model, or even possibly point to
new physics.  This is why we are excited about improving our
measurements of $D$, $i$ and $M$ for the near-extreme Kerr hole
GRS~1915+105 (see Sec.\ 6.2).

\begin{deluxetable}{llll}
\tablewidth{0pt}
\tablecaption{Spin Results to Date for Eight Black Holes\tablenotemark{a}}
\tablehead{\colhead{} & \colhead{Source} & \colhead{Spin $a_*$}& \colhead{Reference}}

\startdata
1&   GRS 1915+105&    $>0.98$&            \citealt{mcc+2006} \\
2&   LMC X--1&        $0.92_{-0.07}^{+0.05}$&   \citealt{gou+2009} \\
4&   M33 X--7&        $0.84\pm0.05$&            \citealt{liu+2008,liu+2010} \\ 
3&   4U 1543--47&     $0.80\pm0.05$&            \citealt{sha+2006} \\
5&   GRO J1655--40&   $0.70\pm0.05$&            \citealt{sha+2006} \\
6&   XTE J1550--564&  $0.34_{-0.28}^{+0.20}$&   \citealt{ste+2010b} \\
7&   LMC X--3&        $<0.3$\tablenotemark{b}&  \citealt{ddb+2006} \\
8&   A0620--00&       $0.12\pm0.18$&            \citealt{gou+2010} \\
\enddata

\tablenotetext{a}{Errors are quoted at the 68\% level of confidence.}
\tablenotetext{b}{Provisional result pending improved measurements of $M$ and $i$.}

\end{deluxetable}

Error estimates are primitive for the first four spin results
published in 2006 (Table~1).  In recent work on the other four black
holes, the principal sources of observational error, as well as the
uncertainties in the key model parameters (e.g., the viscosity
parameter), have been treated in detail.  In particular, in our most
recent paper on XTE J1550--564, we exhaustively explored many
different sources of error (see Table~3 and Appendix A in
\citealt{ste+2010b}).  The upshot of the work to date is that in every
case the uncertainty in $a_*$ is completely dominated by the errors in
the three key dynamical parameters that we input when fitting the
X-ray spectral data.  As discussed in Section~3, these parameters are
the distance $D$, the black hole mass $M$, and the inclination of the
inner disk $i$ (which we assume is aligned with the orbital angular
momentum vector of the binary; \citealt{lil+2009}).  In order to
determine the error in $a_*$ due to the combined uncertainties in $D$,
$M$ and $i$, we perform Monte Carlo simulations assuming that these
parameters are normally and independently distributed (e.g., see
\citealt{gou+2009}).  Note that the errors introduced by our use of
the NT model (Sec.~5), which are not considered here, are
significantly smaller than the observational errors (see Table~7 and
Sec.~4 in \citealt{kul+2010}).

We now discuss the results for four black holes in some detail.  We
first consider the persistent source M33 X--7.  We then turn to GRS
1915+105, the prototype microquasar \citep{mir+1994}, which hosts a
near-extreme Kerr hole.  Finally, we consider the microquasars
A0620--00 and XTE J1550--564, contrasting their behavior with that of
GRS 1915+105.

\subsection{M33 X-7: The First Eclipsing Black Hole}

A long observation of the galaxy M33 with the {\it Chandra} X-ray
Observatory led to the discovery of a black hole that is eclipsed by
its companion star \citep{pie+2006}.  We made a detailed follow-up
dynamical study of the optical counterpart of M33~X-7, the first such
study of a black hole binary beyond the environs of the Milky Way.  We
determined a precise mass for the black hole, $M = 15.65 \pm
1.45$~\msun~\citep{oro+2007}, as well as the mass of its exceptional
companion star ($\approx70$~\msun).  As we discuss in Orosz et al., it
is difficult to understand the origin of this system -- a massive
black hole in a 3.5-day orbit, separated by only 42 solar radii from
its supergiant companion (see Fig.\ 1).  Recently, a consistent
evolutionary model has been proposed that accounts for all the key
properties of the system \citep{val+2010}.  It assumes that
M33 X-7 started as a primary of $\sim95$~\msun~and a secondary of
$\sim30$~\msun, with an orbital period that is close to its present
3.5-day value.

Using as input our precise values for the black hole mass and orbital
inclination angle, and the well-established distance of M33, we fitted
15 {\it Chandra} and {\it XMM-Newton} X-ray spectra and obtained a
precise value for the spin of the black hole primary, $a_*=0.84 \pm
0.05$ \citep{liu+2008,liu+2010}.  Remarkably, given that an
(uncharged) astrophysical black hole is described by just its mass and
spin, this result yields a complete description of an asteroid-size
object at a distance of 2.74 million light-years (840 kpc\footnote{1
parsec (pc) = 3.26 light-years.}).

What is the origin of the spin of M33 X-7?  Was the black hole born
with its present spin, or was it torqued up gradually via the
accretion flow supplied by its companion?  In order to achieve a spin
of $a_* = 0.84$ via disk accretion, an initially non-spinning black
hole must accrete $5.7 M_\odot$ from its donor star \citep{kin+1999}
in becoming the $M = 15.65 M_\odot$ black hole that we observe today.
However, to transfer this much mass even in the case of
Eddington-limited accretion requires $>17$ million
years\footnote{$\dot{M}_{\rm Edd} \equiv L_{\rm Edd}/{\eta}c^2$, where
$L_{\rm Edd} = 1.3{\times}10^{39}{M/10M_{\odot}}$~erg s$^{-1}$ and the
efficiency $\eta$ increases from 5.7\% to 13.3\% as $a_*$ increases
from 0 to 0.84 \citep{sha+1983}.}, whereas the massive companion star,
and hence its host system, can not be older than about 2--3 million
years \citep{oro+2007}.  Thus, it appears that the spin of M33 X-7
must be chiefly natal -- i.e., the event horizon trapped much of the
angular momentum of the collapsing stellar core -- a conclusion that
has been reached for two other stellar-mass black holes
\citep{mcc+2006,sha+2006}.  (However, see \citealt{mor+2008} on
hypercritical accretion).

\subsection{GRS 1915+105: A Near-Extreme Kerr Black Hole}

GRS 1915+105 has unique and striking properties that sharply
distinguish it from the 50 or so known binaries that are believed to
contain a stellar-mass black hole \citep{mcc+2006,oze+2010}.  It is the
most reliable source of relativistic radio jets in the Milky Way and
is the prototype of the microquasars \citep{mir+1994}.  It frequently
displays extraordinary X-ray variability that is not mimicked by any
other black hole system.  The properties of its high-frequency X-ray
oscillations are equally extraordinary \citep{rem+2006}.  Among the 17
transient black hole systems, GRS~1915+105 is unique in having
remained active for more than a decade since its discovery in 1992.
The system has an orbital period of 30.8 days, and it is the largest
of the black hole binary systems (Fig.\ 1).

The pc-scale radio jets of GRS 1915+105, with apparent velocities
greater than the speed of light (superluminal motion), are the
analogue of the kpc-scale jets that have long been observed for
quasars.  The source episodically ejects material at relativistic
speeds, which can easily be tracked for weeks at centimeter
wavelengths as clouds of plasma moving outward on the plane of the sky
\citep{mir+1994,fen+1999}.  Based on a kinematic model, the jet
velocity for a plausible distance of $D\sim11$~kpc is $v_{\rm J}/c >
0.9$, and the inclination of the jet from our line of sight is $i_{\rm
J} \approx 65^{\circ}$.

Based on the analysis of X-ray spectral data for reasonable estimates
of $D$, $M$ and $i$, we discovered that GRS 1915+105 contains a
near-extreme Kerr hole with $a_*>0.98$ \citep{mcc+2006}.  However, the
current estimates of both $D$ and $M$ are poor: The distance is
uncertain by a factor of $\approx 2$ (Fig.\ 5$b$), and the mass is
uncertain by $\approx30$\% ($M = 14.4 \pm 4.4$~\msun;
\citealt{gre+2001,har+2004}).

Remarkably, the extraordinarily high spin of GRS 1915+105 and other
properties of the source allow one -- for this source only -- to place
tight constraints on the allowable values of $M$ and $D$, which must
lie within the triangular region shown in Figure~5$a$.  Our spin model
constrains the black hole's mass and distance to lie to the right of
the slanted line (99\% confidence level) because the model implies
values of $a_*>1$ to the left of the line.  (Our model is only valid
for $a_*<0.999$).  Distances $>12$~kpc are ruled out by the kinematic
model of the radio jets (Fender et al.\ 1999).  The lower bound on $M$
is an estimate and is based on work in progress: We (Danny Steeghs et
al.)  are in the act of obtaining near-infrared spectroscopic data at
ESO's VLT Observatory that we fully expect will improve the
measurement of $M$ by at least a factor of two.  When this result is
in hand, we will have $M$ and $D$ constrained to lie within the small
shaded triangle, thereby constraining the distance to lie within the
range 9.5--12~kpc.

As noted above, the distance is also highly uncertain.  Furthermore,
all measurements to date are model-dependent estimates with large,
systematic uncertainties that are difficult to assess (Fig.\ 5$b$).
With Mark Reid, we are in the act of obtaining a model-independent
trigonometric distance with an uncertainty of 10\% via a parallax
measurement (the gold-standard method in astronomy) using the Very
Long Baseline Array (VLBA), a worldwide array of radio telescopes.  We
have made successful observations at four epochs in 2008--2010 and
anticipate that observations at several additional epochs will be
required to reach our goal.  Two hypothetical and possible outcomes of
these VLBA observations are indicated in Figure~5$a$: VLBA2 would
confirm our spin model and VLBA1 would rule it out.  

\subsection{A0620--00 and XTE J1550--564: Two Schwarzschild-Like Black
  Holes}

The host systems of these two black holes are quite small (Fig.\ 1).
The optical companion in A0620--00, which has an orbital period of
only 0.3~days, is a star somewhat cooler than the Sun with about half
its size and mass.  During its yearlong X-ray outburst in 1975--1976,
this nearby transient ($D\approx1$~kpc) became the brightest celestial
X-ray source ever observed (apart from the Sun).  For several days,
the flux at Earth from this source was greater than the combined flux
of all of the hundreds of other X-ray binaries in our galaxy.  During
this period, A0620--00 was also a bright transient radio source, which
was observed with the early radio telescopes of the day.  A reanalysis
of these data by \citet{kuu+1999} indicates that multiple jet
ejections occurred.  The authors find that, like GRS 1915+105, the
radio source was extended on parsec scales, and they infer a
relativistic expansion velocity.

The cool companion star in XTE J1550--564 also has a mass about half
that of the Sun, although its radius is about twice as great (Fig.\
1); the orbital period of the system is 1.5~days \citep{oro+2011}.
During its principal 1998--1999 outburst cycle, this transient source
produced one of the most remarkable X-ray flare events ever observed
for a black hole binary.  For $\approx1$~day, the source intensity
rose fourfold and the flux in the dominant nonthermal component of
emission rose by the same factor.  Then, just as quickly, the source
intensity declined to its pre-outburst level \citep{sob+2000}.  Four
days later, small-scale superluminal radio jets were observed
\citep{han+2009}; their separation angle and relative velocity link
them to the impulsive X-ray flare.  The subsequent detection of
pc-scale radio jets in 2000 led to the discovery of relativistic {\it
X-ray} jets \citep{cor+2002}.  All of the available evidence strongly
indicates that these pc-scale X-ray and radio jets are associated with
the powerful X-ray flare.

Using our recently-determined estimates of $D$, $M$ and $i$ for
A0620--00 and XTE J1550--564 \citep{can+2010,oro+2011}, we fitted the
X-ray spectra of these black holes and determined their spins
(Table~1).  Figure~6 shows a pair of fitted spectra for the latter
source. The spectrum in the left panel is completely dominated by the
thermal component and is therefore ideal for the determination of
spin.  Now, however, using our improved methodologies
\citep{ste+2009a, ste+2010b}, we are able to obtain useful and
consistent values of spin as well for spectra that have a strong
Compton component of emission, like the one shown in the right panel
of Figure~6.  The origin of this component is widely attributed to
Compton upscattering of the soft disk photons by coronal electrons
(see Sec.~7).

The CF spins of both A0620--00 and XTE J1550-564 are quite low:
$a_*\approx 0.1$ and $a_*\approx0.3$, respectively.  The corresponding
nominal radii of their ISCOs are $5.7M$ and $5.0M$, which differ only
modestly from the Schwarzschild value of $6M$.  The low spins of these
two microquasars challenge the long-standing and widely-held belief
that there is a strong connection between black hole spin and
relativistic jets \citep[][~hereafter BZ]{bla+1977}.  If relativistic
jets are powered by black hole spin, then theory predicts that jet
power will increase dramatically with increasing $a_*$
\citep{tch+2010}.  For low spins, the black hole contributes very
little power; in fact, for $a_*<0.4$, the accretion disk apparently
provides more power than the black hole \citep{mck+2005}.

Given the low spins of XTE J1550--564 and A0620--00, it would appear
that their episodic jets are driven largely by their accretion disks.
One well-known candidate mechanism is the centrifugally driven outflow
of matter from a disk described by \citet[][~hereafter BP]{bla+1982}.
A useful comparison of the operational regimes of BP and BZ is given
by \citet{gar+2010}.  They show that BP is always viable, but that BZ
is a more likely mechanism for the most rapidly rotating sources, such
as the extreme-Kerr black hole GRS~1915+105 (see Sec.\ 6.2).  In
closing, we note that a statistical study by \citet{fen+2010}, which
is based on data of uneven quality, found no evidence that black hole
spin powers jets.

\section{The Fe K$\alpha$ Reflection Method}

In the Fe K$\alpha$ method, one determines $r_{\rm ISCO}$ by modeling
the profile of the broad and skewed iron line, which is formed in the
inner disk by Doppler effects, light bending, relativistic beaming,
and gravitational redshift \citep{fab+2000,rey+2003,mil+2007}.  Of
central importance is the effect of the gravitational redshift on the
red wing of the line.  This wing extends to very low energies for a
rapidly rotating black hole ($a_*\sim1$) because in this case gas can
orbit near the event horizon, deep in the potential well of the black
hole.  Relative to the CF method, measuring the extent of the red wing
in order to infer $a_*$ is hindered by the relative faintness of the
signal.  However, the Fe K$\alpha$ method has the virtues that it is
independent of $M$ and $D$, while the blue wing of the line even
allows an estimate if $i$.  As noted in Section~1, this method, while
applicable to both classes of black holes, is presently the only
viable approach to measuring the spins of supermassive black holes.

For stellar-mass black holes, in addition to the thermal disk
component of emission (which is central to the CF method), a
higher-energy power-law component of emission is always observed
(e.g., see Fig.~6).  For supermassive black holes in active galactic
nuclei (AGN), this power-law component is dominant, and it is thought
to be produced by inverse Compton scattering of soft thermal photons
in a hot ($kT\sim100$~keV) corona \citep{rey+2003,don+2007}.
Meanwhile, the disks of both stellar-mass and supermassive black holes
are too cool ($kT\sim1$ and 0.01~keV, respectively) to produce the
observed Fe K$\alpha$ emission line.  Rather, this dominant line, and
a host of other lines, are generated via X-ray fluorescence as a
result of irradiation of the disk by the hard, coronal power-law
component.  The complex spectrum so generated is referred to as a
``reflection spectrum.''  In order to determine the spin using the
Fe-line method, one must model the reflection spectrum in detail
\citep{ros+2005,ros+2007}.

The spins of several stellar-mass black holes have been measured using the
Fe K$\alpha$ method.  An early suggestion of high spins for two black
holes was made by \citet{mil+2002,mil+2004} and preliminary results
for a total of eight stellar-mass black holes are given in
\citet{mil+2009}.  Other important papers on the spins of stellar-mass
black holes include \citet{rei+2008,rei+2009}.  For a review, see
\citet{mil+2007}.  Very recently, we teamed up with Fe~K$\alpha$
experts to measure the spin of XTE J1550--564 \citep{ste+2010b}. The
spin estimate obtained using the Fe K$\alpha$ method is $a_* =
0.55_{-0.22}^{+0.15}$, which is quite consistent with the CF value
(see Table~1).

The spins of several supermassive black holes have been reported,
which range from $a_* \approx 0.6$ to $>0.98$
\citep{bre+2006,fab+2009,sch+2009,min+2009}.  By far, the most well
studied of these is the Seyfert 1 galaxy MCG--6--30--15 (for
background, see \citealt{rey+2003}).  The 6.4 keV Fe K$\alpha$ line of
this AGN is extremely broad and skewed.  \citet{bre+2006} and
\citet{min+2009} show that the red wing extends downward to below
4~keV and conclude that $a_*>0.98$.

\section{Conclusion}

We have discussed the only two established classes of black holes,
stellar-mass and supermassive\footnote{There is evidence for a class
of intermediate-mass ($100-10^5$) black holes \citep{mco+2000}.
However, to date there existence remains uncertain because no direct
and confirming measurement of mass has been made.}, and the two
principal approaches to measuring their spins, the continuum-fitting
and Fe-K$\alpha$ methods.  Spin measurements for eight stellar-mass
black holes are presented, and these data are used to argue that the
high spin of M33 X-7 is natal, and that at least some relativistic
black-hole jets are powered by their accretion disks, not the spin of
the black hole.

Two aspects of this work excite us greatly. 

First, by measuring a black hole's spin, after earlier measuring its
mass, we are able to {\it completely} characterize the intrinsic
properties of each of the black holes we study.  The No Hair Theorem
states that a macroscopic black hole, regardless of how massive it may
be, is described by just two parameters: $M$ and $a_*$\footnote{In
principle it could also have an electric charge, but astrophysical
black holes are unlikely to have enough charge to be dynamically
important.}. But is the No Hair Theorem really true? The only way we
will answer this question is by first measuring $M$ and $a_*$ for a
good sample of black holes, and then testing whether the Kerr metric
corresponding to these values of $M$ and $a_*$ is completely
consistent with all observables that are sensitive to the space-time
near the black hole. In a sense, we have attempted the first test of
the No Hair Theorem by measuring the spin of XTE J1550-564 by two
independent methods, the continuum fitting method and the Fe K$\alpha$
method (Sec. 7), and finding agreement. However, the errors in the two
measurements are currently rather large, and we do not yet understand
all systematic sources of error, so it would be premature to claim
success.  But this example provides a taste of how astrophysics can
contribute to deep questions in physics.

The other aspect that excites us is all the areas of astrophysics that
our work ties to, e.g., the connections that are beginning to be made
between measurements of spin and the phenomenology and theory of
relativistic jets (Sec.~6.3), and the processes that lead to black
hole formation (Sec.~6.1).  We hope to see spin data used to help
constrain models of gamma-ray bursts, black hole formation, black-hole
binary evolution, high- and low-frequency X-ray oscillations, black
hole X-ray states and state transitions, models of X-ray coronae, etc.

These two strong motivations stimulate us to continue firming up the
measurements of black hole spin, with the goal of amassing a good
sample of a total of 12--15 measurements during the next several
years.

We conclude by noting that it is reasonable to expect LIGO, LISA and
other gravitational-wave observatories to provide us with intimate
knowledge concerning black holes.  However, these breakthroughs are
still some years in the future whereas astrophysical techniques are
providing information on black holes today.  Also, gravitational-wave
facilities are unlikely to help us understand MHD accretion flows in
strong fields, or the origin of relativistic jets, or the formation of
relativistically-broadened Fe lines and high-frequency quasi-periodic
oscillations, etc., phenomena that are now routinely observed for
black holes.  In short, observations of {\it accreting} black holes
show us uniquely how a black hole interacts with its environment.

There is no straight path to unlocking the mysteries of black holes,
probing the extreme physical conditions they generate, and
understanding their importance to astrophysics and cosmology.  It
behooves us to explore widely because it is the synergistic
exploration of all paths that will enlighten us.  Therefore, it is
important to maintain balance between gravitational-wave and
electromagnetic studies of black holes.

\acknowledgments

J.E.M. acknowledges support from NASA grant NNX08AJ55G and the
Smithsonian Endowment Funds.  R.N. was supported in part by NSF grant
AST-0805832 and NASA grant NNX08AH32G.  We thank Laura Brenneman and
Jon Miller for helpful comments on the Fe-line method.

\clearpage


\clearpage

\begin{figure}[ht]
\centering
\includegraphics[scale=0.90,angle=0]{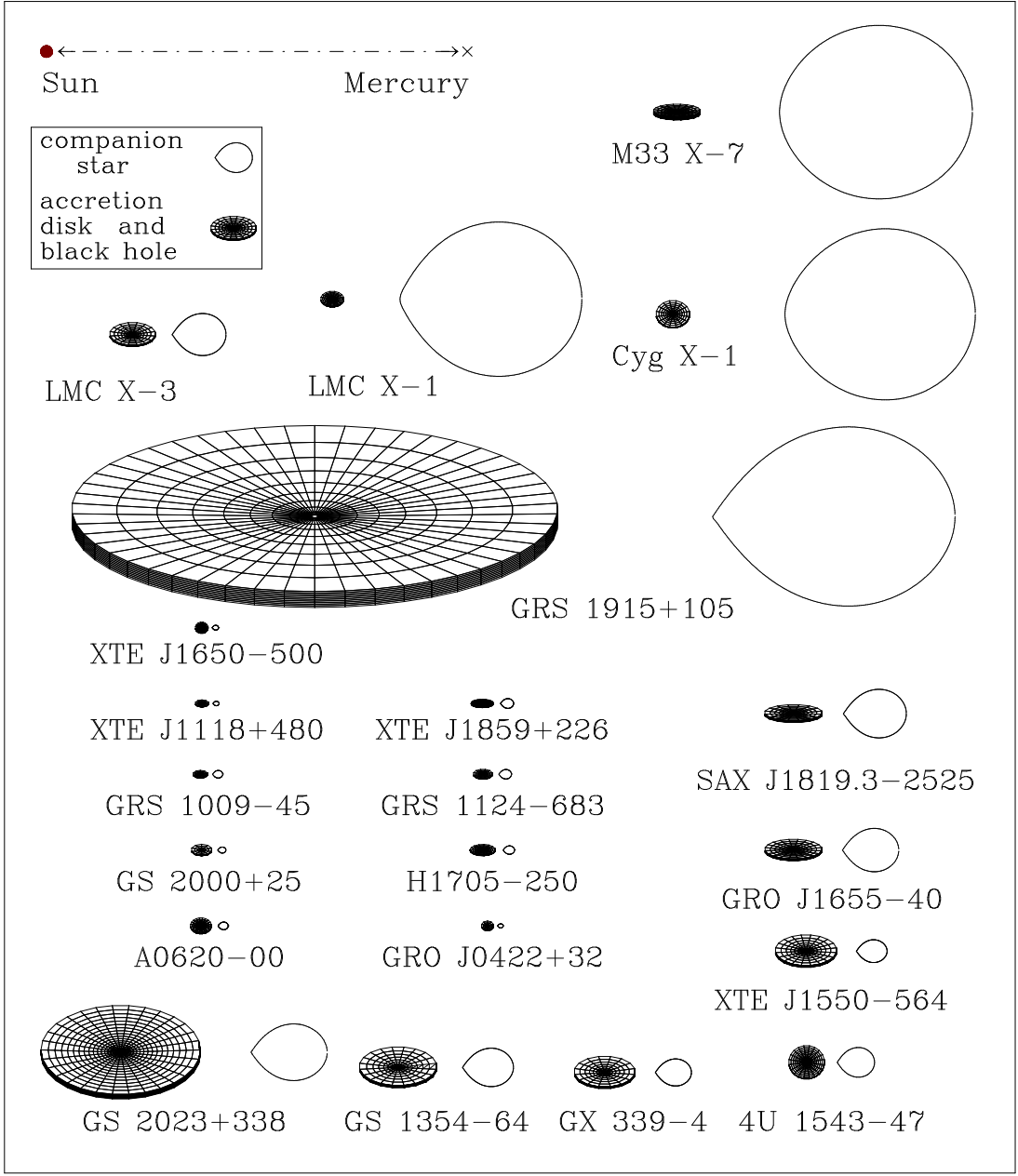}
\caption{Scale drawings of 21 black hole binaries.  The size of the
Sun and the Sun-Mercury distance (0.4 AU) are indicated at the top.
The systems range in size from the giant GRS~1915+105 with an orbital
period of 30.8~days to tiny XTE~J1118+480 with an orbital period of
0.2 days.  The shapes of the tidally distorted stars are accurately
rendered, and the black hole is located at the center of the accretion
disk (see key in inset).  The inclination of the binary to our line of
sight is indicated by the tilt of the accretion disk; an inclination
angle of $i=0^{\circ}$ corresponds to a system whose accretion disk
lies in the plane of the sky and is viewed face on (e.g., $i=
21^{\circ}$ for 4U 1543--47 and $i=75^{\circ}$ for SAX J1819.3-2525).}
\end{figure}

\clearpage

\begin{figure}[ht]
\centering
\plotone{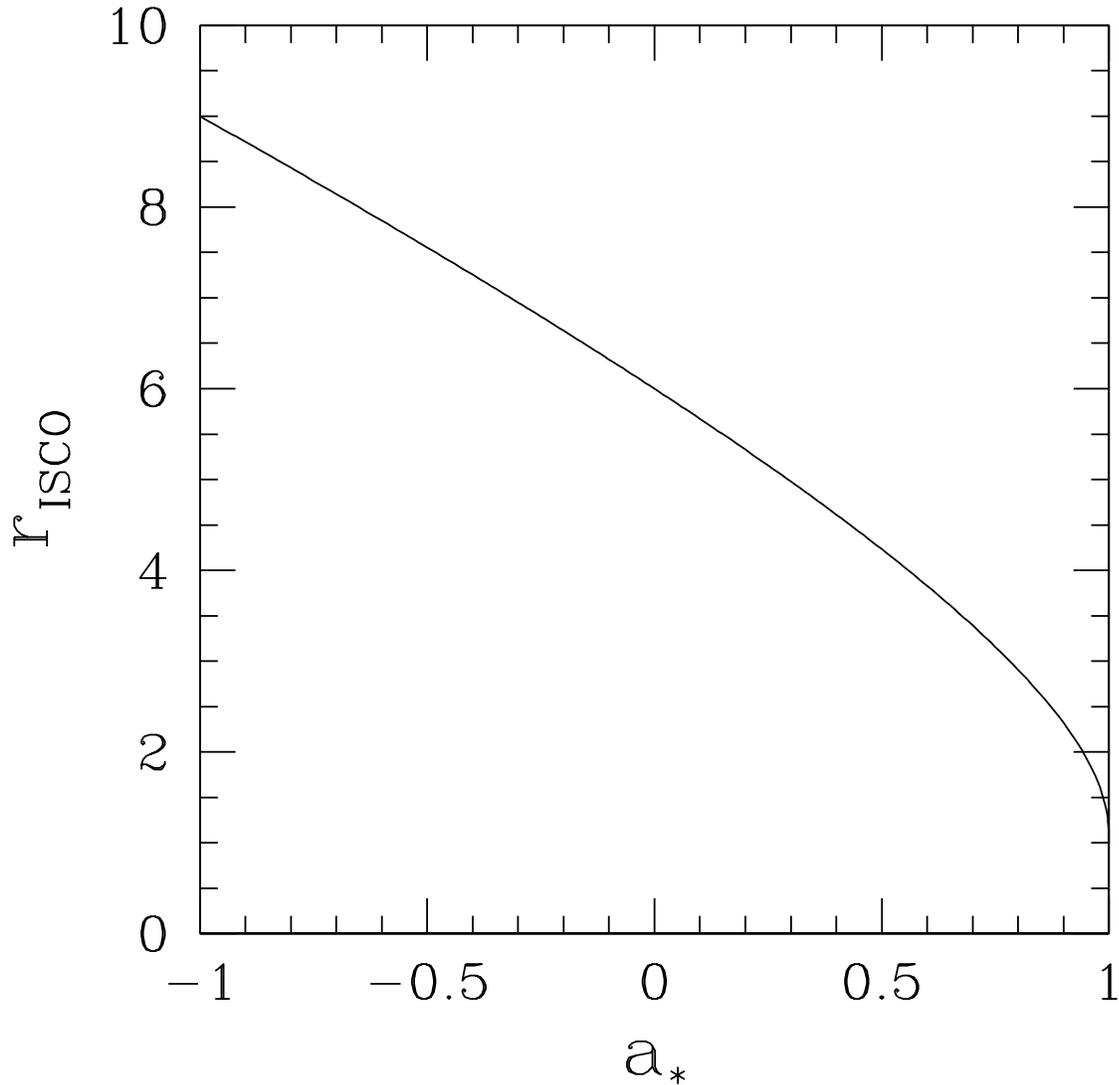}
\caption{Radius of the ISCO in units of $GM/c^2$ versus the black hole
spin parameter.  Negative values of $a_*$ correspond to retrograde
motion, with the black hole spinning in the opposite sense of the
disk.  Stellar black holes are expected to have prograde spins
($a_*>0$) as a consequence of their formation in a binary system,
whereas the spins of supermassive black holes, which are conditioned
by galaxy merger events, may be either prograde or retrograde (e.g.,
\citealt{gar+2010}).}
\end{figure}

\clearpage

\begin{figure}[ht]
\centering
\plotone{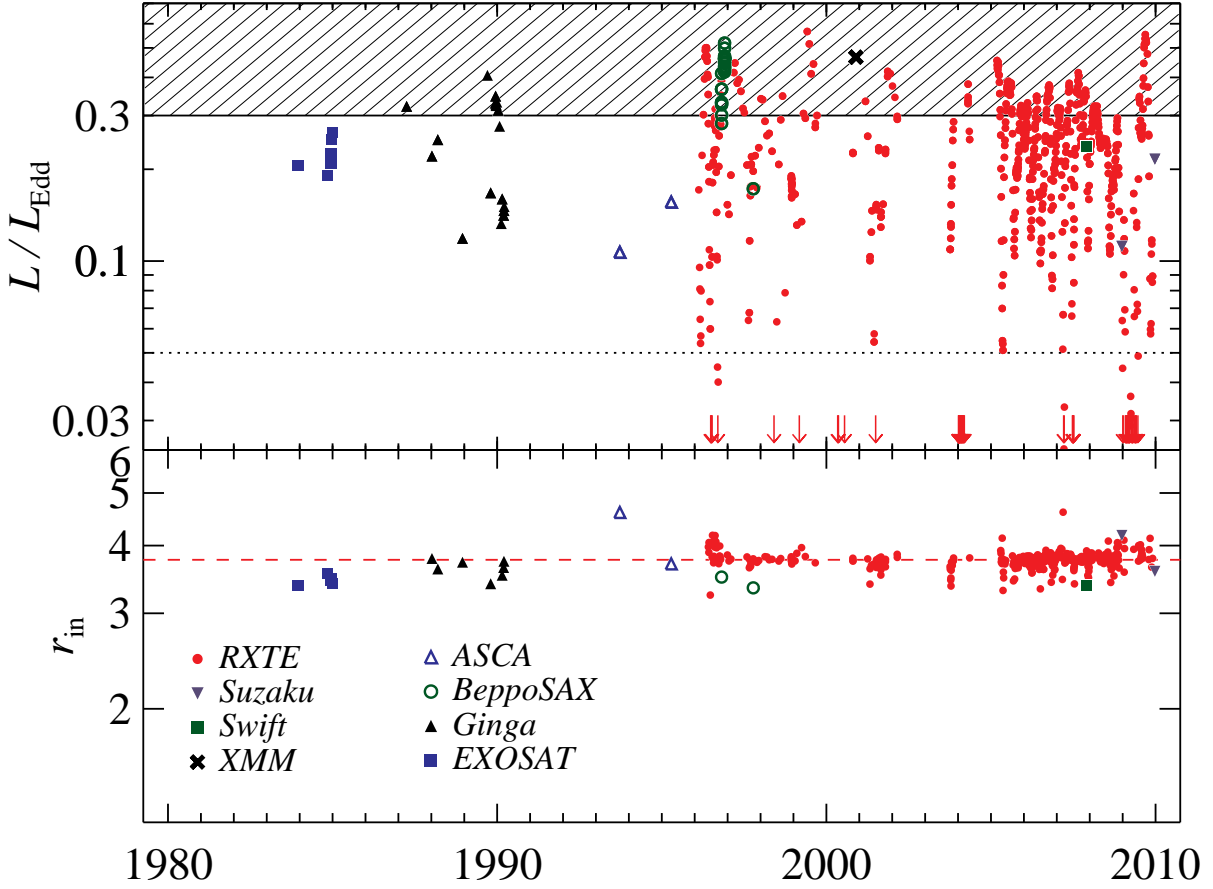}
\caption{$(top)$ Accretion-disk luminosity in Eddington-scaled units
(for $M=10$~\msun) versus time for all the 766 spectra considered in
the study of LMC X-3 by \citet{ste+2010a}.  The downward arrows show
{\it RXTE} data which are off scale.  Data in the unshaded region
satisfy our thin-disk selection criterion $L/L_{\rm Edd} < 0.3$ (Sec.\
3).  The dotted line indicates the lower luminosity threshold
(5\%~$L/L_{\rm Edd}$) set to avoid confusion with strongly Comptonized
data.  $(bottom)$ Fitted values of the inner disk radius $r_{\rm in}
\equiv R_{\rm in}/(GM/c^2)$ are shown for thin-disk data in the top
panel that meet the selection criteria of the study (a total of 411
spectra).  Despite large variations in luminosity, $r_{\rm in}$
remains constant to within a few percent over time.  The median value
for just the 391 selected {\it RXTE} spectra is shown as a red dashed
line.}
\end{figure}

\clearpage

\begin{figure}[ht]
\centering
\plotone{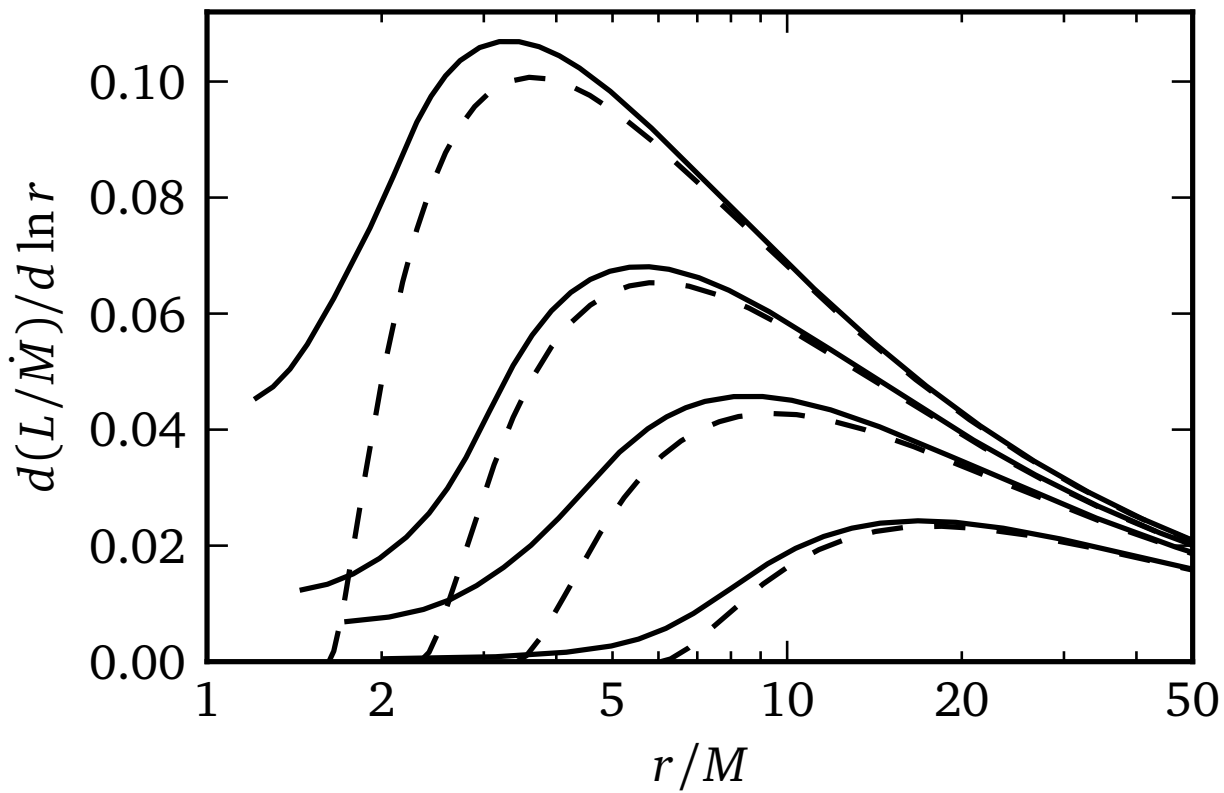}
\caption{Luminosity profiles from GRMHD simulations (solid lines)
compared with those from the \citeauthor{nov+1973} model
(dashed lines) for $a_*=0$, $0.7$, $0.9$ and $0.98$ (bottom to
top). The ISCO is located at the radius where the NT disk luminosity
goes to zero.}
\end{figure}

\clearpage

\begin{figure}[ht]
\centering
\plotone{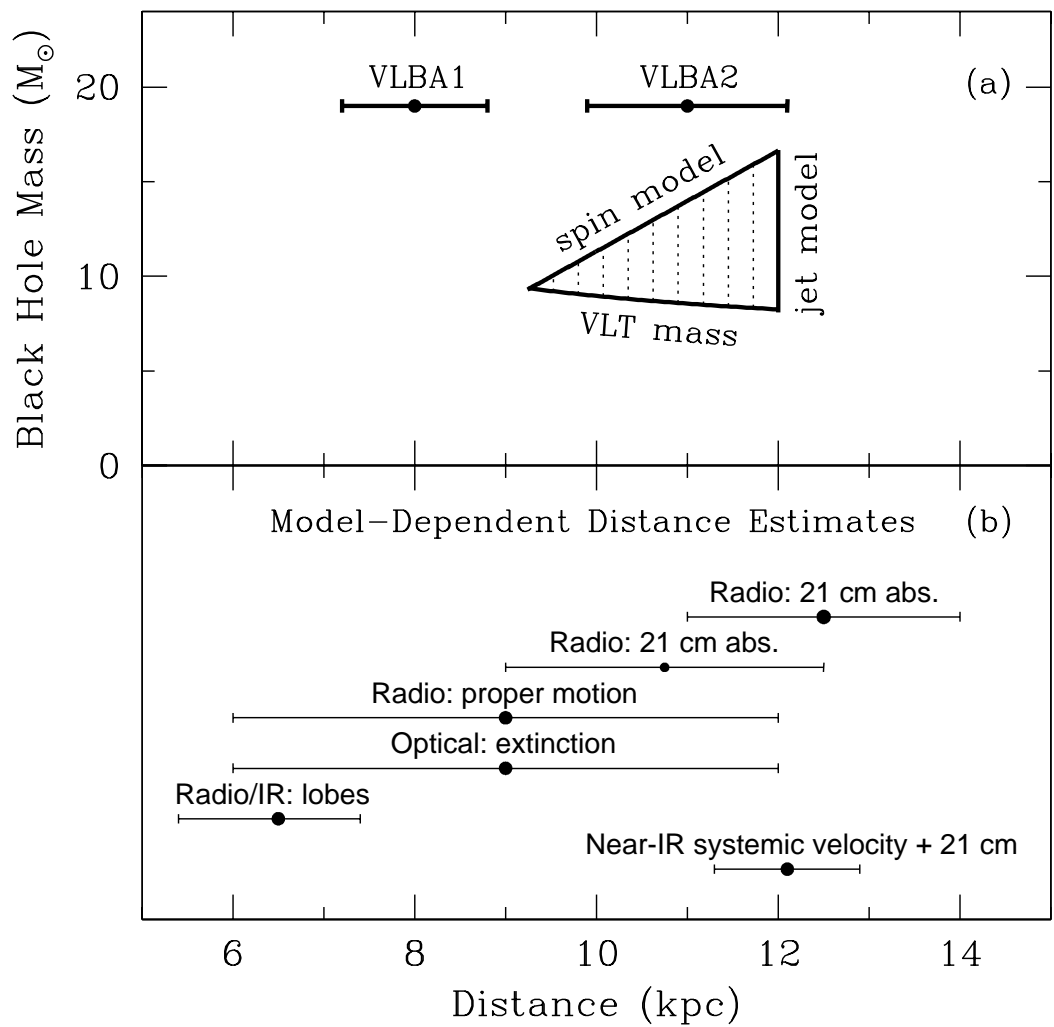}
\caption{($a$) Allowed values of black hole mass and distance for
GRS~1915+105 fall within the shaded triangular region (see text).
($b$) Six estimates of the distance to GRS~1915+105 are shown.  They
range from below 7 to above 12 kpc.  We are working toward a 10\%
trigonometric distance.  Two hypothetical and possible outcomes of our
VLBA observations, labeled VLBA1 and VLBA2, are indicated at the top
of panel $a$.  For references on distance estimates, see Figure~18 in
\citet{mcc+2006}.}
\end{figure}

\clearpage
\begin{figure}[ht]
\centering
\includegraphics[scale=0.70,angle=90]{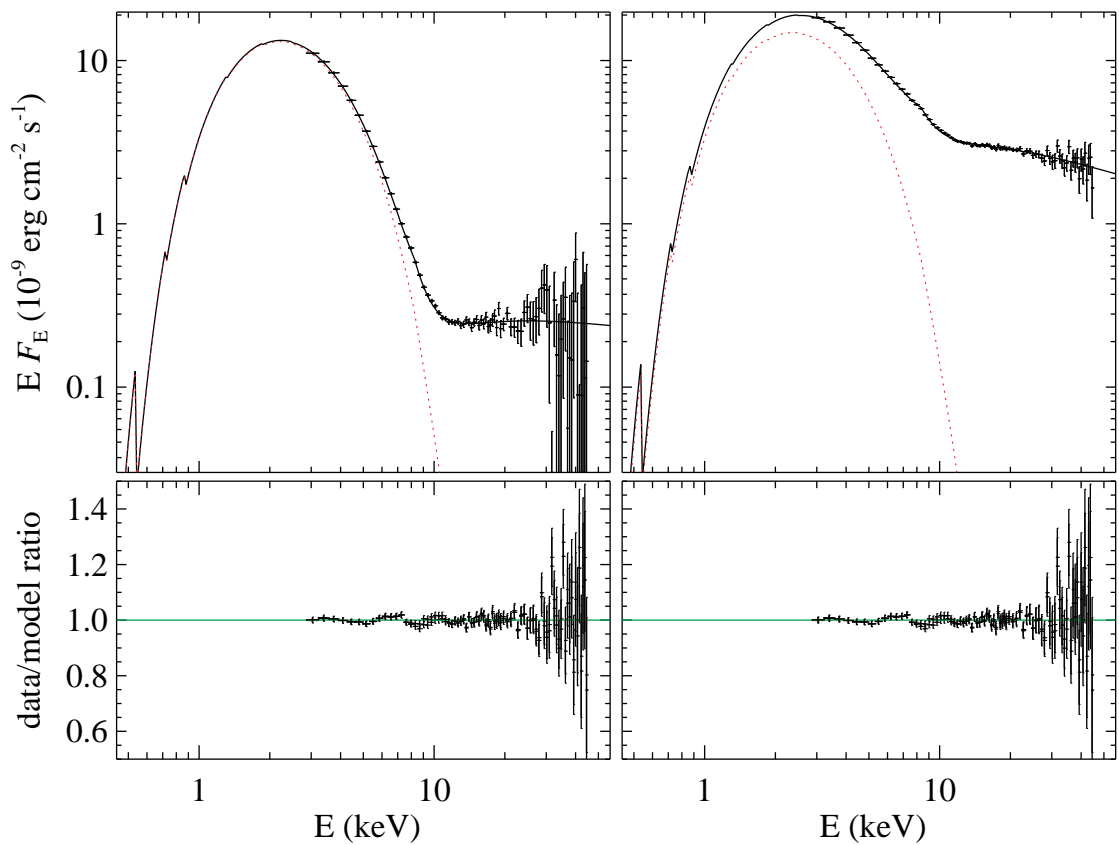}
\caption{Model fits for a pair of spectra of XTE~J1550--564.  $(left)$
A spectrum with a strongly dominant thermal component; shown are the
data, the fit to the data and the fitted thermal component. $(right)$
A strongly Comptonized spectrum.  Note the intensity of the power-law
component relative to its intensity in the left panel.  For details,
see Figure~4 and text in \citet{ste+2010b}.}
\end{figure}

\end{document}